\documentclass[%
 reprint,
 amsmath,amssymb,
 aps,
 prl,
]{revtex4-1}

\usepackage{dcolumn}
\usepackage{graphicx}
\usepackage{bm}
\usepackage{color}
\usepackage{multirow}

\begin{document}

\title{Impurity State and Variable Range Hopping Conduction in Graphene}
\author{Sang-Zi Liang}
\author{Jorge O. Sofo}
\affiliation{Department of Physics and Materials Research Institute, The Pennsylvania State University, University Park, Pennsylvania 16802, USA}

\begin{abstract}
The variable range hopping theory, as formulated for exponentially localized impurity states, does not necessarily apply in the case of graphene with covalently attached impurities. We analyze the localization of impurity states in graphene using the nearest-neighbor, tight-binding model of an adatom-graphene system with Green's function perturbation methods. The amplitude of the impurity state wave function is determined to decay as a power law with exponents depending on sublattice, direction, and the impurity species. We revisit the variable range hopping theory in view of this result and find that the conductivity depends as a power law of the temperature with an exponent related to the localization of the wave function. We show that this temperature dependence is in agreement with available experimental results.
\end{abstract}
\maketitle

Chemical functionalization of graphene has been proposed as one of the most promising methods to modify its transport properties. The attachment of covalently bonded atoms or functional groups opens the possibility to design devices \cite{ribas_patterning_2011,shen_dispersion_2011}, open band gaps \cite{sofo_graphane_2007, nair_fluorographene_2010}, and control the excellent transport properties of its massless Dirac fermions \cite{novoselov_two-dimensional_2005}. Because graphene is essentially a two-dimensional electronic system, even small amounts of this functional group can produce radical changes into its transport properties. The temperature dependence of the conductivity of chemically functionalized graphene shows a peculiar behavior that strongly suggests the importance of disorder. This phenomenon is very general and has been observed with hydrogen \cite{elias_control_2009, matis_giant_2012}, fluorine \cite{withers_electron_2010, hong_colossal_2011}, oxygen \cite{gomez-navarro_electronic_2007, leconte_damaging_2010, moser_magnetotransport_2010}, and metals \cite{li_electron_2011}. In order to relate the experimental measurements with microscopic properties, this anomalous temperature dependence has been analyzed with a variable range hopping (VRH) theory as formulated for semiconductors with exponentially localized impurity states \cite{mott_conduction_1969, mott_metal-insulator_1978}. As a consequence, the estimated localization length does not seem to correlate with any reasonable characteristic length in these systems. In the case of dilute fluorinated graphene, the localization length is obtained to be 56~nm \cite{hong_colossal_2011}, while it is estimated to be 90~nm in lightly silver-coated graphene \cite{li_electron_2011}. The VRH theory, as originally formulated by Mott, is not applicable when the localization of the impurity states is not exponential. For the case in hand, it has been extensively discussed that the impurities form resonant states and the low temperature conductivity as a function of carrier concentration has been theoretically determined in good agreement with experimental results \cite{robinson_adsorbate-limited_2008, wehling_resonant_2010, yuan_modeling_2010, ferreira_unified_2011}. Here, we determine the power law decay of these impurity states and show that the exponent depends on the resonant energy and approaches asymptotically the case of vacancies \cite{pereira_disorder_2006, sherafati_impurity_2011}. The exponent is also anisotropic, displaying strong dependence on the sublattice and on the direction.
We use our findings to reformulate the VRH theory, under the assumption that there is a regime of temperature and density where the jump between these impurity states is incoherent, and determine a general behavior that explains the measured temperature dependence of the conductivity in these systems. The use of this reformulated theory provides a method to determine the localization characteristics of the impurity states in graphene.

\begin{figure*}
\scalebox{0.16}{\includegraphics*{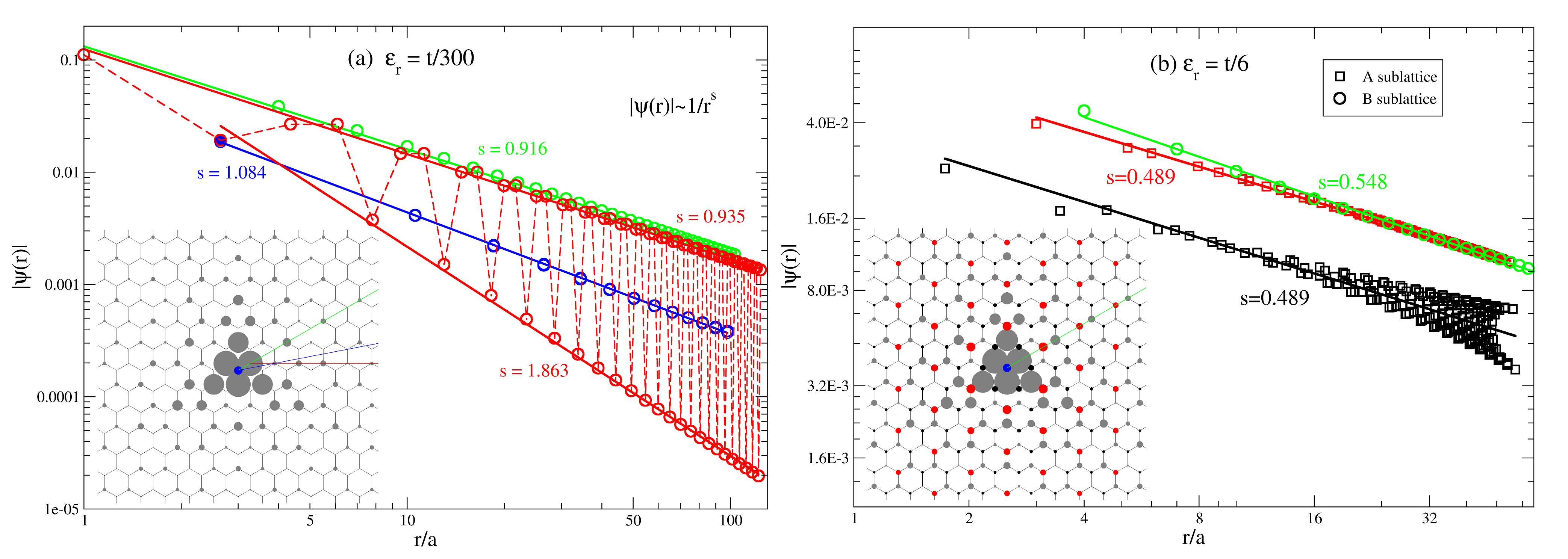}}
\caption{\label{ABsite}Amplitude of the resonance state at different resonance energies. (a) $\epsilon_{r}=t/300$. The amplitudes on the \textit{A} sublattices vanish. (b) $\epsilon_{r}=t/6$. The insets show the amplitudes represented by the radii of circles on the graphene honeycomb lattice. In both plots a circle symbol represents an amplitude on a \textit{B} site, which is in the direction indicated by a line with the same color in the insets. A square symbol represents an amplitude on an \textit{A} site, which has a circle drawn with the same color in the inset. The center dot (blue) is where the adatom is attached. $a$ is the nearest neighbor carbon-carbon distance.}
\end{figure*}

We start by studying the system of a single impurity atom on graphene with a tight-binding Hamiltonian of a localized $p_z$-orbital basis set $\left|i\right>$ for the $\pi$ band of the pristine graphene and $\left|\textup{ad}\right>$ for the adatom,
\begin{equation}
H=H_{0}+H'
\end{equation}
\begin{equation}
H_{0}=-t\sum_{\textup{n.n.}}\left|i\right>\left<j\right|
\end{equation}
\begin{equation}
H'=\epsilon_{\textup{ad}}\left|\textup{ad}\right>\left<\textup{ad}\right|+V_{\textup{ad}}\left(\left|0\right>\left<\textup{ad}\right|+\textup{H. c.}\right),
\end{equation}
where $t$ is the hopping energy between nearest-neighbor carbon atoms ($\approx2.8$ eV), $\epsilon_{\textup{ad}}$ is the site energy of the adatom, and $V_{\textup{ad}}$ is the hopping energy between the adatom and the carbon atom to which it is attached ($\left|0\right>$). This model is often used to study the adatom-graphene system \cite{wehling_local_2007, robinson_adsorbate-limited_2008, pereira_modeling_2008}. Treating $H'$ as a perturbation, the $T$ matrix is given by
\begin{align}
T ={}& H'+H'G_{0}H'+H'G_{0}H'G_{0}H'+\cdots \\ \nonumber
={}& \left|\textup{ad}\right>\frac{V_{\textup{ad}}}{1-V_{\textup{ad}}^{2}G_{0}^{00}G_{0}^{\textup{ad}}}\left<0\right|+\textup{H. c.} \\ \nonumber
  & +\left|\textup{ad}\right>\frac{V_{\textup{ad}}^{2}G_{0}^{00}}{1-V_{\textup{ad}}^{2}G_{0}^{00}G_{0}^{\textup{ad}}}\left<\textup{ad}\right|+\left|0\right>\frac{V_{\textup{ad}}^{2}G_{0}^{\textup{ad}}}{1-V_{\textup{ad}}^{2}G_{0}^{00}G_{0}^{\textup{ad}}}\left<0\right|
\end{align}
where $G_{0}$ is the Green's function of $H_{0}$, $G_{0}^{00}\equiv\left<0\right|G_{0}\left|0\right>$, and $G_{0}^{\textup{ad}}\equiv\left<\textup{ad}\right|G_{0}\left|\textup{ad}\right>=(E-\epsilon_{\textup{ad}})^{-1}$. The perturbed eigenstate in the band continuum is given by the Lippman-Schwinger equation
\begin{align} \label{lippman-schwinger}
\left|\psi(E)\right>={}&\left|\psi_{0}(E)\right>+G_{0}(E)T(E)\left|\psi_{0}(E)\right> \\ \nonumber
={}&\left|\psi_{0}(E)\right>+\frac{V_{\textup{ad}}^{2}G_{0}^{\textup{ad}}\left<0\mid\psi_{0}(E)\right>}{1-V_{\textup{ad}}^{2}G_{0}^{00}(E)G_{0}^{\textup{ad}}(E)}G_{0}(E)\left|0\right>.
\end{align} 
For a certain energy $E=\epsilon_{r}$ that satisfies the resonance condition
\begin{equation}  \label{resonance_condition}
\textup{Re}[1-V_{\textup{ad}}^{2}G_{0}^{00}(\epsilon_{r})G_{0}^{\textup{ad}}(\epsilon_{r})]=0,
\end{equation}
the second term in Eq.~(\ref{lippman-schwinger}) would be significantly enhanced and $\left|\psi_{0}(E)\right>$ can be ignored near the impurity site, which gives
\begin{equation} \label{prop_gf}
\left<i\mid\psi(\epsilon_{r})\right>\propto \left<i\mid G_{0}(\epsilon_{r})\mid 0\right>.
\end{equation}
Similar results have been obtained by other authors \cite{huang_resonant_2009, sherafati_impurity_2011, huang_density_2010}. This resonance state will have larger amplitude near the adatom, but it never decays to zero for large distance because of the contribution of the Bloch function $\left|\psi_{0}(E)\right>$. Also, note that this expression only depends on the resonance energy $\epsilon_{r}$, which means the values of $\epsilon_{\textup{ad}}$ and $V_{\textup{ad}}$ in the Hamiltonian only contribute to determine $\epsilon_{r}$ through Eq.~(\ref{resonance_condition}), and we can study the effects of different adatom species by varying $\epsilon_{r}$.

The decay of the wave function of the impurity state can be studied by investigating the lattice Green's functions (GFs) $\left<i\mid G_{0}(\epsilon_{r})\mid 0\right>$, which are determined mostly from contributions from the two Dirac points ($\mathbf{K}$, $\mathbf{K'}$) in the Brillouin zone when they are evaluated at energies near zero. Integrating around the two Dirac points rather than the whole BZ and assuming a completely linear band, the GFs have been calculated and given in terms of Hankel functions \cite{sherafati_rkky_2011, wang_modeling_2006, nanda_electronic_2012} (labeled \textit{A} for sites in the same sublattice as the impurity and \textit{B} for sites in the opposite sublattice)
\begin{equation} \label{GFapproxA}
\left<\mathbf{r},\textit{A}\mid G_{0}(E)\mid 0\right>=-i\beta\frac{A_{c}E}{4v_{F}^{2}}H_{0}^{(1)}\left(\frac{Er}{v_{F}}\right)
\end{equation}
\begin{equation} \label{GFapproxB}
\left<\mathbf{r},\textit{B}\mid G_{0}(E)\mid 0\right>=-\alpha\frac{A_{c}E}{4v_{F}^{2}}H_{1}^{(1)}\left(\frac{Er}{v_{F}}\right),
\end{equation}
where $A_{c}$ is the area of a unit cell in graphene and $v_{F}$ is the Fermi velocity. The amplitude of the Hankel functions $H_{0}^{(1)}$ and $H_{1}^{(1)}$ decay isotropically, but the GFs also depend on the prefactors
\begin{equation}
\alpha\equiv e^{-i\pi/3}(e^{i{\mathbf{K}\cdot\mathbf{r}-\theta_{r}}}-e^{i{\mathbf{K'}\cdot\mathbf{r}+\theta_{r}}})
\end{equation}
\begin{equation}
\beta\equiv e^{i\mathbf{K}\cdot\mathbf{r}}+e^{i\mathbf{K'}\cdot\mathbf{r}},
\end{equation}
where $\theta_{r}=\tan^{-1}(r_{y}/r_{x})$ when the x axis is taken to be along $\mathbf{K}'-\mathbf{K}$. The form of the argument of the Hankel function makes two types of approximations feasible. For an impurity with a small $\epsilon_{\textup{ad}}$ or a large $V_{\textup{ad}}$ (e.g., a vacancy), the resonance energy $\epsilon_{r}$ solved from Eq.~(\ref{resonance_condition}) will be small, which means we can do small argument expansion to the Hankel function. This gives a resonance state that has zero amplitude on the \textit{A} sublattice sites and decays as $r^{-1}$ for the \textit{B} sites \cite{nanda_electronic_2012,pereira_disorder_2006}. On the other hand, when $\epsilon_{r}$ is not vanishingly small, we are more interested in the long-range decaying behavior, and it is necessary to do large argument expansion, which gives
\begin{equation} \label{largeexpan}
\left|H_{\nu}^{(1)}\left(\frac{Er}{v_{F}}\right)\right|=\left(\frac{2v_{F}}{\pi Er}\right)^{\frac{1}{2}}\left(1+\frac{\left(4\nu^{2}-1\right)v_{F}}{8Er}+\cdots\right).
\end{equation}
Given enough distance, both the \textit{A}-site and the \textit{B}-site amplitude will fall off primarily as $r^{-0.5}$.

The decay behavior is further elucidated by evaluating the GFs directly. The method used here to obtain the GFs for the honeycomb lattice follows a calculation for square lattice \cite{berciu_efficient_2010}, in which the lattice GFs are calculated from larger to smaller distances from the impurity. This is done to avoid a diverging term that originates from numerical instabilities and also satisfies the GF equation of motion \cite{horiguchi_lattice_1972, berciu_computing_2009}. 

\begin{figure}
\scalebox{0.57}{\includegraphics*{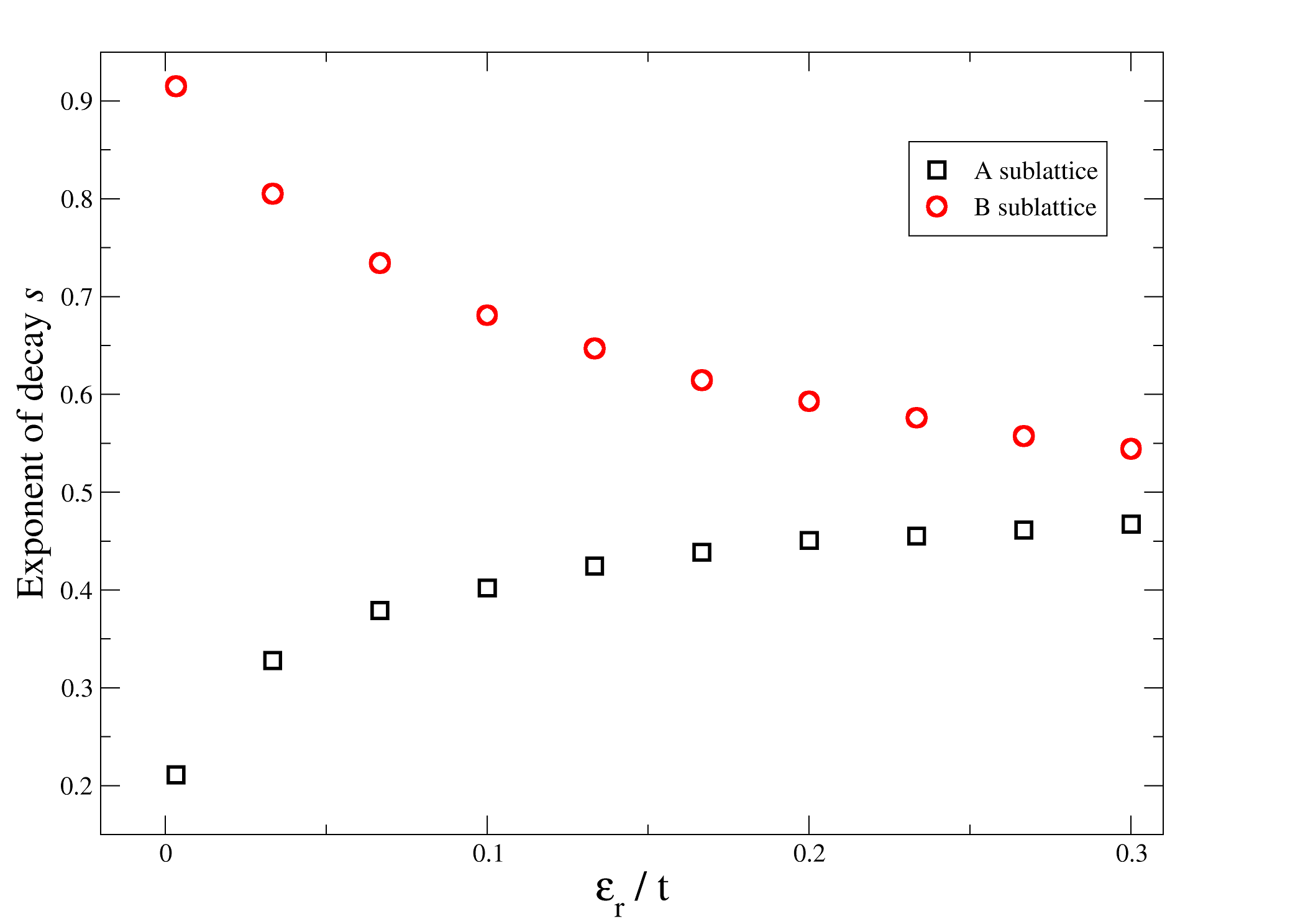}}
\caption{\label{decay_slope}The characteristic decay exponents of the two sublattice sites vs. the resonance energy. For the \textit{A} sites, the exponent is taken from the sites that forms a triangular lattice with the impurity site, similar to the top line in Fig.~\ref{ABsite}(b). For the \textit{B} sites, the exponent is taken from sites in the armchair direction, similar to the green line in Fig.~\ref{ABsite}.}
\end{figure}

The calculated GFs (amplitudes of the resonance state) are plotted in Fig.~\ref{ABsite}, where in the insets the amplitude of the GF on a certain site is represented by the radius of the circle that is drawn on that site, and the results mostly confirm the approximations. Firstly, the amplitude of the resonance state wave function depend on the resonance energy $\epsilon_{r}$, and the energy dependent behaviors of the two sublattice sites are drastically different. This is clear from Fig.~\ref{ABsite}(a) ($\epsilon_{r}=t/300$) and Fig.~\ref{ABsite}(b) ($\epsilon_{r}=t/6$). At low $\epsilon_{r}$, the resonance state is almost exclusively on the \textit{B} sublattice sites. The amplitudes on the \textit{A} sites increase quickly while the \textit{B} sites stay relatively the same with increasing $\epsilon_{r}$, and the two sublattice sites have comparable amplitudes at $\epsilon_{r}=t/6$. Secondly, the wave function amplitude decays with power law
\begin{equation} \label{power-law}
\left|\psi(r)\right|=\frac{\psi_{0}}{r^{s}},
\end{equation}
although the exponent $s$ depends on resonance energy, sublattice, and direction. For the \textit{A} sites [Fig.~\ref{ABsite}(b)], the sites that form a triangular lattice with the impurity site (marked with red) have larger amplitudes than the other sites (marked with black) when they are approximately the same distance to the impurity. This can be explained with the prefactor $\beta$ in Eq.~(\ref{GFapproxA}), which evaluates to 2 for the former group of sites and -1 for the latter. The two groups give almost perfect linear fits in a log-log plot close to the impurity with essentially the same decay exponent~$s$, while the second group of sites deviates from the linear fit at larger distance. For the \textit{B} sites, the decay is anisotropical and power laws can be seen in many directions when $\epsilon_{r}$ is small (Fig.~\ref{ABsite}(a)). This behavior has been obtained with approximations to the GFs previously by Nanda \textit{et al}. \cite{nanda_electronic_2012}, and our calculation confirms their result. At higher $\epsilon_{r}$, the decay in the \textit{B} sites in the armchair direction is the slowest (and thus contributes the most when calculating overlap) and still obeys very good power laws, while the other data sets start to deviate. Deviations from power laws in both the \textit{A} sites and the \textit{B} sites are not explained in the approximated GFs, and they happen at a smaller distance for larger $\epsilon_{r}$. Therefore, it is possibly a result of nonzero energy and contributions from k points other than the two Dirac points. The decay exponents for the \textit{A} sites and the armchair direction of the \textit{B} sites are plotted in Fig.~\ref{decay_slope} with several resonance energies. As predicted by the approximations, the \textit{B} sites' decay exponent is 1 at zero resonance energy \cite{nanda_electronic_2012, pereira_disorder_2006}, and both exponents approach 0.5 with large resonance energy. Finally, we would like to note that the amplitude of the wave function calculated with GFs agrees with evaluation of the GFs with elliptic integrals \cite{horiguchi_lattice_1972}, the results in Ref.~\onlinecite{pereira_modeling_2008} for vacancies ($\epsilon_{r}=0$), as well as the amplitude of the eigenfunction near the impurity ($r\leq 10a$) obtained from direct diagonalization of the Hamiltonian with periodic boundary conditions.

\begin{table*}
\begin{ruledtabular}
\begin{tabular}{c|cc|cc}
& \multicolumn{2}{l|}{Conventional VRH} & \multicolumn{2}{l}{Power Law VRH} \\
\hline
Data Set & $\Omega$ & $T_{0}$\ (K) & $\Omega$ & $\eta$ \\
\hline
Hydrogenated, $V_{g}=0\textup{V}$, Ref.~\onlinecite{elias_control_2009} & 0.9857 & 284 & 0.9930 & 0.714 \\
Hydrogenated, $V_{g}=0\textup{V}$, Ref.~\onlinecite{matis_giant_2012} & 0.9884 & 280 & 0.9975 & 0.524 \\
Hydrogenated, $V_{g}=3\textup{V}$, Ref.~\onlinecite{matis_giant_2012} & 0.9819 & 187 & 0.9981 & 0.459 \\
Hydrogenated, $V_{g}=9\textup{V}$, Ref.~\onlinecite{matis_giant_2012} & 0.9621 & 107 & 0.9928 & 0.382 \\
Fluorinated, $V_{g}=0\textup{V}$, Ref.~\onlinecite{withers_electron_2010} & 0.9843 & 450 & 0.9730 & 0.826 \\
Fluorinated, $V_{g}=0\textup{V}$, Ref.~\onlinecite{hong_colossal_2011} & 0.9990 & 270 & 0.9761 & 0.664 \\
Fluorinated, $n=0.7\times 10^{12}\textup{cm}^{-2}$, Ref.~\onlinecite{hong_colossal_2011} & 0.9985 & 130 & 0.9843 & 0.550 \\
Fluorinated, $n=1.4\times 10^{12}\textup{cm}^{-2}$, Ref.~\onlinecite{hong_colossal_2011} & 0.9889 & 33 & 0.9955 & 0.335 \\
Fluorinated, $n=2.5\times 10^{12}\textup{cm}^{-2}$, Ref.~\onlinecite{hong_colossal_2011} & 0.9599 & 5 & 0.9929 & 0.190 \\
\end{tabular}
\end{ruledtabular}
\caption{\label{expdata}The fitting details of available data sets with both the conventional VRH and Eq.~(\ref{power_law_sigma}). $V_{g}$ is the experimental gate voltage relative to the charge neutrality point (the voltage where the sample exhibits the highest resistance). $n$ is the charge density calculated with the gate voltage and sample specifics. $T_{0}$ is the characteristic temperature in the conventional VRH $\ln\sigma\propto-(T_{0}/T)^{1/3}$ and $\Omega$ is the squared correlation coefficient for the fits on the linearized data.}
\end{table*}

The power-law decay and the Bloch-wave behavior at large distance both indicate that the impurity state in graphene is not nearly as localized as a typical midgap state in a semiconductor, which decays exponentially. The resonance states here are not normalizable, and it would be difficult to define a localization length. In view of the simplicity and the extensive usage of the VRH theory, it is necessary to investigate the impact of a power-law-decaying impurity state on the hopping conductivity result.

Similar to VRH, we will assume that the density of impurities is low enough so that the average distance between impurities is longer than the phase-coherent length and that coherent scattering from multiple centers can be ignored. The derivation of VRH is outlined in Refs.~\onlinecite{mott_conduction_1969, shklovskii_electronic_1984}. By simply replacing the overlap with Eq.~(\ref{power-law}), the hopping probability between two impurity sites $i$ and $j$ can be written as
\begin{equation}
P_{ij}=\gamma_{ij}\frac{1}{r_{ij}^{2s}}\textup{exp}\left(-\frac{\epsilon_{ij}}{kT}\right),
\end{equation}
where $r_{ij}$ and $\epsilon_{ij}$ are the distance and energy difference between the two states, respectively. $\gamma_{ij}$ is a prefactor that comes from the coupling between electrons and phonons. It depends on $\epsilon_{ij}$, and $r_{ij}$ with power laws and is thus ignored in the original VRH \cite{shklovskii_electronic_1984}. Assuming a smooth density of states near the Fermi level ($g(\epsilon_{\textup{F}})$), $\epsilon_{ij}$ and $r_{ij}$ are related by
\begin{equation}
2g(\epsilon_{\textup{F}})\epsilon_{ij}=r_{ij}^{-d},
\end{equation}
where $d$ is the dimension of the system. Taking the occupation of the two states into account, we obtain the conductance between two impurities as \cite{ambegaokar_hopping_1971, shklovskii_electronic_1984}
\begin{equation}
\sigma_{ij}= \frac{e^2}{kT}P_{ij}.
\end{equation}
where $e$ is the charge of an electron. The conductivity of the bulk is assumed to be proportional to $\sigma_{ij}$ between the most conductive pair of impurities, which corresponds to the maximum of $\sigma_{ij}$ when varying $r_{ij}$ or $\epsilon_{ij}$. This leads to a power-law temperature dependence for the conductivity,
\begin{equation} \label{power_law_sigma}
\sigma\propto T^{\eta},\ \textup{with}\ \eta=\frac{2s}{d}+s',
\end{equation}
where $s'$ originates from the prefactor $\gamma_{ij}/T$. For hydrogenic states, it is estimated to be $(\nu-2)/(d+1)$, where $\nu$ is the critical exponent for the size of the percolating cluster \cite{shklovskii_electronic_1984}. In two dimensions, $\nu=1.34$ \cite{dunn_series_1975} and
\begin{equation} \label{sprime}
s'=-0.22
\end{equation}
The analysis that gives this exponent cannot be easily generalized, but we can still assume $s'$ to be a constant that does not depend on the details of the impurities.

Equation.~(\ref{power_law_sigma}) can be used to fit existing experimental data of the systems of hydrogen adatoms \cite{elias_control_2009, matis_giant_2012} and fluorine adatoms on graphene \cite{hong_colossal_2011} and the extracted parameters are shown in Table~\ref{expdata}, along with fits of the original VRH. The fits are done to all the data points presented in the references. Assuming Eq.~(\ref{sprime}), the exponents extracted from the fittings are within the reasonable range that is expected from this theory. In both experiments where the effect of gate voltage is studied, the exponent decreases when the gate voltage moves away from the charge neutrality point, effectively shifting the Fermi energy so that on average impurity states with higher resonance energies participate in the conduction. This is consistent with the behavior of the \textit{B} sites' decay exponent (Fig.~\ref{decay_slope}) while the amplitudes on the \textit{A} sites are too small to be relevant in the range of the experimental gate voltage.

The conventional VRH and Eq.~(\ref{power_law_sigma}) have very similar curvatures in this temperature range and all the data can fit both equations fairly well. Comparing the correlation coefficients ($\Omega$) for the linear fits, the data for fluorinated graphene at low gate voltage fit the original VRH better, while the power-law dependence describes all the other data sets better. The currently available data cannot convincingly exclude either equation as the conduction mechanism, especially considering the VRH requires low temperature but all the data go up to room temperature. We expect more continuous and accurate experimental data at low temperature to prove our proposal of a power-law temperature dependence.

In summary, we have shown that the impurity state in graphene is a resonance state in the band continuum and it is localized only as power-law functions with exponents generally below 1. This means that the VRH theory which assumes exponential localization is not directly applicable to disordered graphene. Replacing the overlap term in VRH, a theory for the temperature dependence of conductivity is derived which fits the existing experimental data. However, since the states are largely delocalized, the hopping picture of conduction may not be the most appropriate approach to model the transport properties of these systems. Further investigation into this problem is needed to develop a theory that includes both the impurity states and the extended unperturbed states.

\begin{acknowledgments}
We are grateful for useful discussions with Prof. Jun Zhu and supported in part by the Materials Simulation Center, a Penn-State Center for Nanoscale Science (MRSEC-NSF Grant No. DMR-0820404) and MRI facility, and by the Research Computing and Cyberinfrastructure Group from ITS-Penn State.
\end{acknowledgments}

\end{document}